\documentstyle[aps]{revtex}

\input epsf.sty

\begin{document}
\draft


\newcommand{\PL}{\Gamma^{\leftarrow}}
\newcommand{\PR}{\Gamma^{\rightarrow}}

\title{General technique of calculating drift velocity and
       diffusion coefficient\\ 
       in arbitrary periodic systems}
\author{Zbigniew Koza}
\address{Institute of Theoretical Physics, University of
Wroc{\l}aw, pl. Maxa Borna 9, PL-50204 Wroc{\l}aw,
Poland\thanks{Permanent address; e-mail: zkoza@ift.uni.wroc.pl.}
}
\address{Institut f\"ur Festk\"orperforschung, Forschungszentrum
J\"ulich GmbH, D-52425 J\"ulich, Germany}

\date{\today}

\maketitle

\begin{abstract}
We develop a practical method of computing the stationary drift
velocity $V$ and the diffusion coefficient $D$ of a particle (or a few
particles) in a periodic system with arbitrary transition rates. We
solve this problem both in a physically relevant continuous-time
approach as well as for models with discrete-time kinetics, which are
often used in computer simulations.  We show that both approaches
yield the same value of the drift, but the difference between the
diffusion coefficients obtained in each of them equals $\case{1}{2}
V^2$. Generalization to spaces of arbitrary dimension and several
applications of the method are also presented.
\end{abstract}



\section{Introduction}

Investigation of diffusive transport is of highest importance in many
areas of physics and related sciences. The most fundamental
characteristics of diffusion is provided by two quantities -- the
diffusion coefficient $D$ and the drift velocity $V$. The knowledge of
the latter is especially important in studies of non-equilibrium
phenomena, where usually $V\neq0$. This includes, among others,
diffusion in disordered media subject to an external field
\cite{Haus-Kehr,Bouchaud,EE,Derrida98,KehrJ,Driven,Katz} or the so
called molecular ``pumps'' and ``motors''
\cite{MotorsA,MotorsB,MotorsC}, responsible 
for transport of various chemicals in biological cells.  However, no
universal {\em and} practical theoretical method of determining $V$
and $D$ for arbitrary systems has been developed, and although many
solutions for some particular physical models have been proposed, in
more complicated cases one often has to resort to approximations or
numerical methods.

The first, to our knowledge, attempt towards determining $V$ and $D$
in an arbitrary system was made by Derrida \cite{Der83JSP}, who
considered a one dimensional, periodic lattice of a period $L$ and
arbitrary hopping rates between nearest-neighbor sites. He managed to
give exact expressions for the velocity and the diffusion constant as
functions of the hopping rates. He then extended his method
\cite{Der83PRB} to a periodic $d$-dimensional system, but his solution
turns out to be rather complicated. An alternative method of
calculating $V$ was also developed by Kehr {\em et al.}
\cite{KehrJ}. Of the two quantities, $V$ and $D$, the latter
is of course much harder to find. A surprisingly simple formula for
one-dimensional systems with transition rates satisfying the detailed
balance condition was derived by Lyo and Richards
\cite{Lyo}, and recently independently reinvented by Kutner
\cite{KutnerPHA} and Wichmann \cite{Wichmann},
\begin{equation}
 \label{Kutner}
 D = \left( 
         \frac{1}{L^2}
         \sum_{j=0}^{L-1} \frac{1}{P^{\rm eq}_j \Gamma^{\rightarrow}_j}
     \right)^{-1}, 
 \end{equation}
where $L$ is the period of the lattice, $P^{\rm eq}_j$ denote the
equilibrium site occupancy probabilities, and $\Gamma^{\rightarrow}_j$
are transition rates from site $j$ to $j+1$. However, this equation is
valid only when the random walker hops between nearest-neighbor sites.
A method of determining $D$ in arbitrary periodic systems satisfying
the detailed balance condition was then proposed by Braun and Sholl
\cite{Braun}.  Description of several other techniques, developed for
some special types of random environments, can be found in review
articles \cite{Haus-Kehr,Bouchaud,EE,Derrida98,Kehr98}.

All the above-mentioned methods are based on the following scheme:
\begin{enumerate}
\item Reduce the infinite system to
a single elementary cell with periodic boundary conditions.
\item Write down the master equation.
\item  Using it calculate the steady-state properties of this
system; in particular, find the steady-state site occupation
probabilities \cite{Der83JSP,Der83PRB,KutnerPHA,Wichmann,Braun} or
propagators\cite{Aslangul}. 
\item Given these quantities, find $V$ and $D$.
\end{enumerate}
%
The third step
is here critical and can be carried out explicitly only for relatively
small systems or for models possessing some special properties, e.g,
one-dimensional lattices with jumps restricted to the nearest-neighbor
sites. The main goal of this paper is to develop a simple and, at the
same time, general method of calculating $V$ and $D$ explicitly
without performing detailed analysis of the steady state. In our
approach the third step reads
\begin{enumerate}
\item[3'.] Find the matrix $\Lambda(k)$ representing 
the Fourier transform of the master equation. 
\end{enumerate}
which is a rather straightforward operation. The technique we propose
is valid for both equilibrium and non-equilibrium systems. It can also
be quite easily implemented in computer-algebra or numerical
programming languages.

Following the first step, our considerations will be restricted to
periodic systems, such as crystals or molecular motors. A general problem
of diffusion in an arbitrary disordered system remains an open
question. One might be tempted to treated it by taking the
limit $L\to\infty$. This method can sometimes provide a hint 
about anomalous type of  diffusion in a given aperiodic system
\cite{Kehr98}. However, except for some simple
models \cite{Bouchaud,AslJSP}, it has not been established whether the
limits $t\to \infty$ and $L\to \infty$ commute and thus, whether this
approach is generally correct.

We will pay a special attention to a one-dimensional system of finite
period $L$ with arbitrary jump rates between any of its sites whose
distance is less than $L$.  This is perhaps the simplest model of a
periodic system where all elements of the transition rate matrix can
take on nonzero values. Its investigation can be therefore carried out
with relatively simple mathematical formalism. A solution to a
general, multi-particle and multi-dimensional problem requires the
same mathematical methods, but the notation would have to be
complex. What is even more important, any periodic system with finite
number of sites in its elementary cell can be mapped on such a
one-dimensional system; geometry of the original problem is relevant
mainly in constructing the matrix $\Lambda(k)$ and can be taken into
account quite easily. Once the explicit form of $\Lambda(k)$ has been
found, one can compute $V$ and $D$ using the methods we derive here
for this simple one-dimensional system.

In our calculations we will use both continuous-time and discrete-time
formalisms. One could expect that since we calculate $V$ and $D$ in a
stationary state, where all quantities are independent of time, both
approaches should yield the same result.  However, as was found by
Derrida \cite{Der83JSP}, the form of the diffusion coefficient as a
function of the transition rates can depend on whether time flows
continuously, or not. This is an important issue, since using
discrete-time models belongs to favorite techniques of computer
physics (in particular the cellular-automata \cite{CA}, and the
exact-enumeration \cite{EE} methods), and one needs to know whether
results obtained in this way are correct.  Derrida \cite{Der83JSP}
showed that in his (grossly simplified) model the relation between the
diffusion constants obtained in the continuous-time ($D$) and
discrete-time ($D^{\rm D}$) formalisms reads $D - D^{\rm D} =
\frac{1}{2}{V}^2$.  We will show that this relation is general and 
will prove it for any periodic system with arbitrary transition rates.

The paper is organized as follows. Using the Fourier transform of the
master equation, see Refs.\ \cite{Haus-Kehr,Kehr98,Kutner77}, in
Section \ref{SecGen1} we analyze a one-dimensional system with
arbitrary period $L$ and transition rates $\Gamma_{jl}$. We find a
simple technique of calculating $V$ and $D$ in terms of the 3 lowest
coefficients of the characteristic polynomial of the
Fourier-transformed transfer-matrix, $\Lambda(k)$.  The most important
properties of its spectrum at $k=0$ has been already described in
Refs.\ \cite{Kutner77,vanKampen}, but only for systems obeying the
detailed balance condition. Therefore, in Sections \ref{SectionCont}
and \ref{SectionDiscrete} we analyze in detail the spectrum of
$\Lambda(k)$ in a general case, assuming that the time is a continuous
or a discrete quantity, respectively. Then, in Section
\ref{Comparison} we compare the two approaches and explain the
different forms of diffusion coefficients derived in each of them.  In
Section \ref{SecTotGen} we generalize our approach to systems in
arbitrary space dimension, and in Section \ref{SecNN} we present a
particularly simple method of calculating $V$ and $D$ for
one-dimensional systems with transitions between the nearest-neighbor
sites. In Section \ref{SecAppl} we derive explicit forms of $V$ and
$D$ in some commonly used models. In particular, one of the examples
studied there explains how to apply our technique to many-body
problems.  Finally, section \ref{SecConcl} is devoted to conclusions.


\section{General case of a periodic 1D system}
\label{SecGen1}

\subsection{Continuous-time 
             formalism}
\label{SectionCont}
Consider a one-dimensional lattice with its sites located at $x_n$,
$n\in{\cal N}$.  At time $t=0$ we put a particle at site $x_0=0$. The
particle can then jump between the lattice sites. Transitions are
assumed to represent a continuous (e.g., a Poisson) Markov process in
time. The (constant in time) transition rate from a site $x_n$ to a
site $x_m$ will be denoted by $\Gamma_{mn}$. 
We assume that the system is periodic in space and denote its
period by $L\ge 1$, 
\begin{equation}
 \label{period}
 \forall_{n,m}  \;\;   \Gamma_{mn} = \Gamma_{m+L,n+L}, 
                \;\;\; x_m - x_n = x_{m+L} - x_{n+L}.
\end{equation} 
Of course transition rates between two different sites cannot be
negative, $\Gamma_{mn} \ge 0$ if $m\neq n$.  For simplicity we also
restrict our considerations to the case where direct transitions
between sites $m$ and $n$ are allowed only if $|m-n| < L$.

Let $P(n,t)$ denote the probability density of finding the particle at
site $x_n$ at time $t$. The evolution of this quantity is governed by
the master equation
\begin{equation}
\label{CMEL}
   \frac{\partial P(n,t)}{\partial t} 
   =  \sum_m 
            \left[
                   \Gamma_{nm} P(m,t) - \Gamma_{mn}P(n,t)
            \right],
\end{equation}
and the initial condition reads
\begin{equation}
   P(n,0) = \delta_{n,0}.
\end{equation}
We can now treat the whole lattice as if it consisted of $L$
sub-lattices and define the probability densities $P_l(j,t)$ of
finding a particle at a given sublattice $l$ at site $n$ at time $t$
as
\begin{equation}
  \label{Theta}
 P_l(n,t) \equiv P(n,t)\delta^L_{n,l},
\end{equation}
where $\delta^L_{n,l}$ is a generalized Kronecker's delta,
$\delta^L_{n,l}=1$ if $n = l\,(\mbox{mod}\,L)$ and $\delta^L_{n,l}= 0$
otherwise.  We can now rewrite (\ref{CMEL}) as a system of $L$ linear
differential equations for $P_l(n,t)$, $l = 0,\ldots,L-1$,
\begin{equation}
\label{CMELL}
   \frac{\partial P_l(n,t)}{\partial t} 
   = \sum_{j=1-L}^{L-1}
          \left[ 
             \Gamma_{l,l+j} P_{l+j}(n+j,t) - \Gamma_{l+j,l}P_l(n,t)
          \right].
\end{equation}
The major advantage of (\ref{CMELL}) as compared with (\ref{CMEL}) is
that its coefficients do not depend on $n$. Therefore we can  
calculate its Fourier transform,
\begin{equation}
\label{FCMELL}
   \frac{\partial \tilde{P}_l(k,t)}{\partial t}  
     = \sum_{j=1-L}^{L-1}
          \left[ 
              \Gamma_{l,l+j} {\rm e}^{ik(x_{l+j}-x_l)} 
                 \tilde{P}_{l+j}(k,t)
             -\Gamma_{l+j,l}
                 \tilde{P}_l(k,t)
          \right],
\end{equation}
where $\tilde{P}_l(k,t) \equiv \sum_n \exp(-ikx_{n}) P_l(n,t)$.

The system of equations (\ref{FCMELL}) can be written in a compact
form using an $L\times L$  matrix $\Lambda_{lj}(k)$, 
\begin{equation}
\label{FCMELLambda}
\frac{\partial \tilde{P}_l(k,t)}{\partial t} = 
  \sum_{j=0}^{L-1} \Lambda_{lj}(k) \tilde{P}_j(k,t),
\end{equation}
where 
\begin{eqnarray}
 \label{DefLambdaK}
 \Lambda_{lj}(k) \equiv
   \left\{
     \begin{array}{ll}
         \Gamma_{lj}    {\rm e}^{ik(x_j-x_l)} 
       + \Gamma_{l,j+L} {\rm e}^{ik(x_{j+L}-x_l)},
       & j<l
       \\[1ex]
         \Gamma_{lj}    {\rm e}^{ik(x_j-x_l)} 
       + \Gamma_{l,j-L} {\rm e}^{ik(x_{j-L}-x_l)},
       & j>l
       \\[1ex]
       -\sum_{m\neq l} \Gamma_{ml},
       & j=l.   
     \end{array}
   \right.
\end{eqnarray}

We will impose only one restriction on the form of transition rates
$\Gamma_{lj}$. We will demand that $\Lambda(0)$ be irreducible (by a
permutation of indices) \cite{Gantmacher,Minc}, i.e., in the
stationary state the particle can be found at any of the sublattices
defined in (\ref{Theta}) with a probability $>0$. Physically this
condition means that in the long-time limit the system does not split
up into several non-interacting subsystems.  Mathematically
irreducibility means that $\Lambda(0)$ has exactly one eigenvector
whose all components are strictly positive. Our approach is thus quite
general --- we do not require that the transition rates should satisfy
the detailed balance condition, the Fourier-transformed transition
rate matrix, $\Lambda(k)$, need not be symmetric or even
diagonalizable at $k=0$, and some of its eigenvalues can be complex.

A general  solution to (\ref{FCMELLambda}) reads \cite{Hartman}
\begin{equation}
\label{FPL_KT}
   \tilde{P}_l(k,t) = \sum_{j=0}^{L-1}
      T_{lj}(k,t)\exp[\lambda_j(k)t] ,\; l= 0,\ldots, L-1, 
\end{equation}
where the coefficients $T_{lj}(k,t)$ are polynomials in $t$ and can be
determined using the initial condition.  The degree of $T_{lj}(k,t)$
is smaller than the multiplicity of $\lambda_j(k)$. We assume that the
eigenvalues $\lambda_j(k)$ are ordered in accordance with the
descending magnitude of their real parts at $k=0$, i.e., $j < l
\Rightarrow \Re(\lambda_j(0)) \ge \Re(\lambda_l(0))$.  Since 
$P(n,t) = \sum_{l=0}^{L-1}P_l(n,t)$, there is $\tilde{P}(k,t) =
\sum_{l=0}^{L-1}\tilde{P}_l(k,t)$, and so (\ref{FPL_KT}) yields
\begin{equation}
\label{FPKT}
\tilde{P}(k,t) = \sum_{l=0}^{L-1}h_{l}(k,t)\exp[\lambda_l(k)t],
\end{equation}
where $h_l(k,t) \equiv \sum_{j=0}^{L-1}T_{jl}(k,t)$ are polynomials in
$t$ (actually $h_l(k,t)$ can depend on time only if 
$\lambda_l(k)$ is degenerated).

In a one-dimensional system the stationary drift velocity $V$ and the
diffusion coefficient $D$ are given by
\begin{eqnarray}
  \label{GeneralDefJ}
  V &=& \lim_{t\to\infty} 
          \frac{\langle x \rangle}{t} 
             \nonumber \\[1ex]
    &=&
        \lim_{t\to\infty} 
         i \frac{
                  \left. \frac{\partial \tilde{P}(k,t)}{\partial k}
                   \right|_{k=0}
                }{t}, 
        \\[1ex]
  \label{GeneralDefD}
  D &=&  \lim_{t\to\infty}
            \frac{\langle x^2 \rangle - \langle x \rangle^2}{2t}
              \nonumber  \\[1ex]
    &=& 
         \lim_{t\to\infty} 
              \frac{
 	            -\left. 
                       \frac{\partial^2\tilde{P}(k,t) }{\partial k^2} 
                     \right|_{k=0}
                    +\left(
                        \left.
                           \frac{\partial \tilde{P}(k,t) }{ \partial k}
                        \right|_{k=0}
                     \right)^2                      
                   }{2t},
\end{eqnarray}
where $\langle f(x) \rangle \equiv \int \! f(x)P(x,t)\,dx$.
Thus, if we could calculate the eigenvalues of $\Lambda(k)$, we would
be able, in principle, to calculate $V$ and $D$. The diagonalization
of ${\Lambda(k)}$, however, is a formidable task feasible only in some
special cases (e.g.\ for small $L$ or when special relations have been
imposed on its elements). Fortunately, as we will see below, if our
aim is restricted only to determining $V$ and $D$ as functions of the
transition rates $\Gamma_{lj}$, we need not calculate explicitly even
a single eigenvalue of the transition rate matrix!

Let $\mu$ be a constant such that $\forall_j \; \mu >
|\Lambda_{jj}(0)|$ and let ${\bf Q}$ denote an auxiliary matrix, ${\bf
Q}_{jl} \equiv \Lambda_{jl}(0) + {\mu \delta_{jl}}$.  Since
$\Lambda(0)$ is irreducible, so is ${\bf Q}$. Moreover, because all
elements of ${\bf Q}$ are nonnegative, we can apply to it the
Frobenius theorem \cite{Gantmacher,Minc} and conclude that ${\bf Q}$
has a positive eigenvalue $q$, which is a simple root of the
characteristic equation, and the moduli of all other eigenvalues are
at most $q$. Because the dominant eigenvalue of an irreducible,
nonnegative matrix lies between the largest and smallest column sums
\cite{Gantmacher,Minc}, and in our case these sums are all equal to $\mu$,
we find that the dominant eigenvalue $q = \mu$.  Since the spectrum of
${\bf Q}$ is shifted, with respect to the spectrum of $ \Lambda(0)$,
by $\mu$, we conclude that for $k=0$ the matrix $\Lambda(k)$ has
exactly one dominating eigenvalue $\lambda_0(0)=0$ and the real parts
of all other eigenvalues are negative,
\begin{equation}
 \label{l0}
   \lambda_0(0) 
   =   0 > \Re(\lambda_1(0))
         \ge ,\ldots, \ge
           \Re(\lambda_{L-1}(0)). 
\end{equation}
Thus, in the limit $t\to \infty$ this single eigenvalue dominates the
sum in r.h.s of (\ref{FPL_KT}) at $k\approx 0$.  In determining the
forms of $V$ and $D$ we can therefore employ a simple 
approximation
\begin{equation}
\label{FPKT0}
  \tilde{P}(k,t) = h_0(k,t)\exp[\lambda_0(k)t].
\end{equation}
 Moreover, since $\sum_n P(n,t) = 1$, there is 
\begin{equation}
  \label{One}
  \forall_{t \ge 0} \;\; \tilde{P}(0,t) = 1 ,
\end{equation}
which, owing to (\ref{l0}), implies 
\begin{equation}
\label{h0}
 h_0(0,t) = 1.
\end{equation}
Upon inserting (\ref{FPKT0}) into (\ref{GeneralDefJ}),
(\ref{GeneralDefD}) and using (\ref{l0}), (\ref{h0}) we conclude that
\begin{eqnarray}
  \label{DefJ1L}
  V &=& i \left. \frac{\partial \lambda_0}{\partial k}
                   \right|_{k=0} , 
        \\[1ex]
  \label{DefD1L}
  D &=&   \left.
             -\frac{1}{2}\frac{\partial^2\lambda_0 }{\partial k^2} 
          \right|_{k=0} .
\end{eqnarray}

Let $W(x)$ denote the characteristic polynomial of the matrix
$\Lambda(k)$. Let $c_i(k)$ denote its coefficients at $x^j$, $j =
0,\ldots, L$. We thus have
\begin{equation}
   \label{W}
   \forall_k \;\;
        W(\lambda_0(k)) = \sum_{j=0}^{L} c_j(k)[\lambda_0(k)]^j = 0.
\end{equation}
On differentiating it with respect to $k$ and using the first part of 
(\ref{l0}) we arrive at
\begin{equation}
  \left. \frac{\partial \lambda_0}{\partial k} \right|_{k=0} =
  -\frac{c_0'}{c_1},
\end{equation}
where we used a short-hand notation $c_j \equiv c_j(0)$ and $c_j'
\equiv \partial c_j/\partial k|_{k=0}$. Note that $c_1\neq 0$, which
follows from (\ref{l0}). Differentiating (\ref{W}) twice yields
\begin{equation}
   \left. 
      \frac{\partial^2 \lambda_0}{\partial k^2} 
   \right|_{k=0} 
 =
   - \frac{c_0'' + 2 c_2 (\lambda_0')^2 + 2 c_1'\lambda_0'}{c_1},
\end{equation}
where $\lambda_0'\equiv \partial\lambda_0(k)/\partial k |_{k=0}$
and $c_0'' \equiv \partial^2 c_0/\partial k^2|_{k=0}$. 
We thus finally arrive at our major result
\begin{eqnarray}
  \label{SolJ1C}
  V &=& -i  \frac{c_0'}{c_1}, 
        \\[1ex]
  \label{SolD1C}
  D &=&  
      \frac{c_0'' - 2c_2 V^2 - 2 i c_1'V}{2 c_1}.    
\end{eqnarray}

Functions $c_j(k)$ can depend on $k$ and the transition rates in a
very complicated way. We were able to find only two general
properties, both following immediately from (\ref{l0}): $c_0 = 0$ and
$c_1 \neq 0$. Explicit forms of $c_0'$, $c_0''$, $c_1$, $c'_1$ and
$c_2$ for some particular models will be given below, in sections
\ref{SecNN} and \ref{SecAppl}.


\subsection{Discrete-time
            formalism}
\label{SectionDiscrete}

The discrete-time formulation of the problem is basically similar to
the continuous one, but there are some major differences, too.  For
simplicity we will use the same notation as in the previous section,
but one should remember that almost all functions employed in the
discrete-time formulation of the problem differ from those we dealt
with in Section \ref{SectionCont}. Where it will be necessary to
compare quantities computed within each approach, we will attach a
superscript ``D'' to the quantity derived in the discrete-time
formalism.
 
In the discrete-time version of the problem the master equation
(expressed in terms of the probabilities $P_l(n,t)$ of finding a
particle at time $t$ at a site $x_n$ belonging to a sublattice $l$)
reads
\begin{equation}
\label{DMEL}
  P_l(n,t+1) = 
  P_l(n,t) + 
  \sum_{j=1-L}^{L-1}
  \left[
          \Gamma_{l,l+j} P_{l+j}(n+j,t) - \Gamma_{l+j,l}  P_l(n,t)  
  \right], 
\end{equation}
where $ l = 0,\ldots, L-1$, and $\Gamma_{lj}$ are dimensionless
probabilities satisfying the usual condition $\forall_{lj}\; 0 \le
\Gamma_{jl} \le 1$.  Note that in the continuous formulation of the
problem $\Gamma_{jl}$ were unbounded from above, dimensional
quantities (of dimension [T$^{-1}$]) and we called them ``transition
rates''.  Last, but not least, in the present approach time $t$
assumes only integer values.

Upon taking the Fourier transform of (\ref{DMEL}) we arrive at
\begin{equation}
\label{FDMEL}
    \tilde{P}_l(k,t+1) = 
    \tilde{P}_l(k,t) + 
    \sum_{j=1-L}^{L-1}
    \left[
            \Gamma_{l,l+j} {\rm e}^{ik(x_{l+j}-x_l)} \tilde{P}_{l+j}(k,t)
           -\Gamma_{l+j,l}\tilde{P}_{l}(k,t)
    \right],
\end{equation}
which can be rewritten using a stochastic matrix $\Lambda^{\rm D}_{lj}(k)$
\begin{equation}
\label{DFCMELLambda}
\tilde{P}_l(k,t+1) = 
  \sum_{j=0}^{L-1} \Lambda^{\rm D}_{lj}(k)  \tilde{P}_j(k,t), 
\end{equation} 
where the only difference between $\Lambda^{\rm D}$ and its
continuous-time counterpart $\Lambda$ lies in their diagonal elements
\begin{equation}
   \label{ldl}
   \Lambda^{\rm D}_{lj}(k) = \Lambda_{lj}(k) + \delta_{lj}.
\end{equation}
Therefore the eigenvalues $\lambda^{\rm D}_l$ of
$\Lambda^{\rm D}$ are  related to the eigenvalues 
$\lambda_l$ of $\Lambda$ through a simple formula
\begin{equation}
  \label{lambdy}
  \lambda^{\rm D}_l (k)= \lambda_l(k) + 1,\; l = 0,\ldots, L-1.
\end{equation}

Applying the theorem of Frobenius \cite{Gantmacher,Minc} to
$\Lambda^{\rm D}(0)$, which is a real, nonnegative matrix, we conclude
that
\begin{equation}
 \label{l0=1}
   \lambda^{\rm D}_0(0) = 1
\end{equation}
and the moduli of all other eigenvalues do not exceed 1. However, in
contrast to the continuous-time formalism, now there can be $s\ge1$
eigenvalues, say $\lambda^D_{j_m}(k)$, $m= 0,\ldots,s-1$, such that
$|\lambda^D_{j_m}(0)| = 1$; for convenience we assume that
$\lambda^D_{j_0} \equiv \lambda^D_{0}$. Although this can happen only
if all diagonal elements of $\Lambda^{\rm D}(0)$ vanish
\cite{Gantmacher,Minc}, we do not exclude this exceptional case from
our considerations (see Ref. \cite{Der83JSP}).  The Frobenius theorem
ensures us that all these dominating eigenvalues are distinct, and so
in the normal Jordan representation of $\Lambda^{\rm D}(k)$ the size
of the corresponding Jordan blocks is 1.  This suffices to assert that
for $k\approx 0$ and $t \to \infty$
\begin{equation}
\label{DFPKT}
\tilde{P}(k,t) = \sum_{m=0}^{s-1}h_{m}(k)[\lambda^{\rm D}_m(k)]^t,
\end{equation}
where $h_m$ are some functions of $k$.

Because equation (\ref{One}) is valid both in continuous-time and
discrete-time formalisms, upon comparing it with (\ref{DFPKT}) we
conclude that except for $h_{0}(k)$ all other coefficients $h_{m}(k)$
in (\ref{DFPKT}) must vanish at $k=0$,
\begin{equation}
 \label{Dh1}
   h_{0}(0) = 1, \;\; h_{m}(0) = 0, \; m = 1,\ldots,s-1.
\end{equation}
Actually, since $t$ is an integer, this is not a trivial statement; a
proof is based on the fact that all dominant eigenvalues $\lambda^{\rm
D}_m$ are distinct roots of the equation $x^s =1$, see Refs.\
\cite{Gantmacher,Minc}.  Upon inserting (\ref{DFPKT}) into
(\ref{GeneralDefJ}) and (\ref{GeneralDefD}) and then using
(\ref{l0=1}) and (\ref{Dh1}) we arrive at
\begin{eqnarray}
  \label{DDefJ1L}
  V^{\rm D} &=& i \left. \frac{\partial \lambda^{\rm D}_0(k)}{\partial k}
                  \right|_{k=0} , 
        \\[1ex]
  \label{DDefD1L}
  D^{\rm D} &=&   \left.
              -\frac{1}{2}
              \left[
                 \frac{\partial^2\lambda^{\rm D}_0(k) }{\partial k^2} - 
                 \left(
                    \frac{\partial\lambda^{\rm D}_0(k) }{\partial k}
                 \right)^2
              \right] 
          \right|_{k=0} - \Psi(0,t),
\end{eqnarray}
where 
\begin{equation}
\label{Psi}
  \Psi(k,t) \equiv \sum_{m=1}^{s-1} 
          \frac{\partial h_{m}}{\partial k}
          \left(
            \frac{\partial \lambda_m^{\rm D}}{\partial k} - 
            \frac{\partial  \lambda_0^{\rm D}}{\partial k}
               \lambda_m^{\rm D}
          \right)
          (\lambda_m^{\rm D})^{t-1}.
\end{equation}
Note that, by definition, $\Psi(k,t) \equiv 0$ if $s=1$, i.e., if
$\lambda^{\rm D}_0(0)$ is the only dominating eigenvalue of the
transition matrix $\Lambda(0)$. Moreover, because the existence of the
diffusion constant is guaranteed by the central limit theorem,
$\Psi(0,t)$ must be independent of time, at least in the limit $t\to
\infty$, for any $s$.  This, in turn, requires that even if $s > 1$
\begin{equation}
  \label{Psi0}
  \Psi(0,t) = 0
\end{equation}
(see the comment under eq.\ (\ref{Dh1})). Therefore, to determine $V$
and $D$ one needs only to investigate the properties of a single
eigenvalue $\lambda^{\rm D}_0(0)$ at $k\approx0$, i.e., eq.\
(\ref{DFPKT}) can be replaced with
\begin{equation}
 \label{DFPKT1}
   \tilde{P}(k,t) = h_{0}(k)[\lambda^{\rm D}_0(k)]^t.
\end{equation}

Finally, using (\ref{DefJ1L}), (\ref{DefD1L}), (\ref{lambdy}), and
(\ref{DDefJ1L}) -- (\ref{Psi0}) we conclude that the drift velocities
and diffusion coefficients obtained in continuous and discrete
formulations of the problem are related to each other by simple,
general formulae
\begin{eqnarray}
 \label{JJ}
   V^{\rm D} &=& V, \\[1ex]
 \label{DDtau=1}
   D^{\rm D} &=& D -\case{1}{2} V^2 .
\end{eqnarray}
Equation (\ref{DDtau=1}) holds, however, only if the interval $\tau$ between
successive jumps in the discrete formulation of the problem equals 1,
which has been assumed in our calculations  for simplicity. For a
general value of  
$\tau$ the relation between $D$ and $D^{\rm D}$ reads
\begin{equation}
 \label{DD}
   D^{\rm D} = D - \case{1}{2} \tau V^2 .
\end{equation}
which has a correct dimensional form (all its terms have a dimension
[${\rm L}^2{\rm T}^{-1}$]).

\subsection{Comparison 
            of the two formalisms}
\label{Comparison}

Comparing equations (\ref{FPKT0}) and (\ref{DFPKT1}) we find that the
main mathematical differences between the continuous- and
discrete-time formalisms are related to different functional forms of
$\tilde{P}(k,t)$ at $k\approx 0$. While in the continuous-time
approach this function is given by an exponential, in the
discrete-time model we deal with a power function.  In particular,
notice that $\partial^2\lambda^t(k)/\partial k^2$ yields a term
proportional to $t(t-1) = t^2-t$. This gives rise to an additional
term linear in $t$, which is responsible for the difference between
(\ref{DDefD1L}) and (\ref{DefD1L}), and hence for the term
$\case{1}{2}\tau V^2$ in (\ref{DD}).

The presence of the time unit $\tau$ in (\ref{DD}) suggests that the
key to understanding the physical reasons for the difference between
$D$ and $D^{\rm D}$ lies in the dimensional analysis. 
In the continuous-time approach the transition rates
$\Gamma_{jl}$ are {\em dimensional} quantities that scale with time as
[T$^{-1}$]. Therefore, multiplying them all by some positive constant
$\alpha\neq1$ corresponds to changing the time unit, and so the
resulting diffusion coefficient $D_{\rm new}$ will 
be equal to the product of $\alpha$ and the original value of the
diffusion coefficient, $D_{\rm old}$. Similarly, $V_{\rm new} = \alpha
V_{\rm old}$. This explains why in the
continuous-time approach we could safely assume $\tau =1$.
However, in the discrete formalism there is no such a simple relation
between the time interval $\tau$ and the jump probabilities
$\Gamma_{jl}$, which are {\em dimensionless}. Suppose, for example,
that at times $t=0, \tau, 2\tau, \ldots$ 
we toss a coin. Changing the frequency of tossing, or  $1/\tau$, will
have no impact on the probability of the coin falling heads up. 
The proper way of taking the continuous-time limit in the
discrete-time model is to assume that 
the jump probabilities $\Gamma_{jl}^{\rm D}$ are related to the
continuous-time transition rates $\Gamma_{jl}$ by a simple formula
$\Gamma_{jl}^{\rm 
D} = \tau\Gamma_{jl}$ and taking the limit $\tau \to 0$. In other
words, to get the continuous-time 
limit, we need to apply the ``alpha-transformation'' to the transition
probabilities with an infinitesimally small value of $\alpha$.
Equation (\ref{DD}) shows that $D^{\rm D}$ is actually a sum of two
terms --- one that scales linearly under the
``$\alpha$-transformation'' ($D$), and the other one which scales
quadratically ($-\case{1}{2}\tau V^2$).  In the limit $\tau \to 0$ the
continuous-time diffusion coefficient $D$ is thus of order $O(\tau)$,
while $-\case{1}{2}\tau V^2$ is of order $O(\tau^2$) and can be
neglected. Consequently, as could be expected, the relative difference
between diffusion coefficients calculated within each approach
vanishes, $(D - D^{\rm D})/D \to 0$ as $\tau \to 0$.

A decrease of the diffusion coefficient in the discrete-time formalism
can be also interpreted as a consequence of the fact that a discrete
process tends to be ``less random'' that a continuous one. This is
clearly seen in a limiting case of a one-dimensional lattice (with
lattice constant $a$ and time unit $\tau$) where all probabilities of
jumping to the left vanish ($\Gamma^{\leftarrow}_{j} = 0$) and all
probabilities of jumping to the right are equal $1$
($\Gamma^{\rightarrow}_{j}=1$).  A discrete process with this choice
of probabilities is completely deterministic. At each time interval
the diffusing particle with probability 1 hops to the right, hence the
diffusion constant vanishes. On the other hand, if we consider a
continuous-time process with $\Gamma^{\leftarrow} = 0$ and
$\Gamma^{\rightarrow}=1$, we can see that the motion of the particle
is now by no means deterministic. Although we know that on average
there will be one jump to the right per unit time, we do not know when
it will actually occur. This uncertainty introduces randomness to the
process, and the corresponding diffusion coefficient equals
$a^2/2\tau$.

Note finally that because the left-hand side of (\ref{DD}) represents
a diffusion coefficient, it cannot be negative, and so 
\begin{equation}
  \label{D>V2}
  D \ge \case{1}{2}\tau V^2.
\end{equation}
This has several
interesting consequences. First, if in a periodic system there is a
stationary drift ($V \neq 0)$, then  the
continuous-time diffusion coefficient $D$ is bounded from below.
 That such a bound exists can indirectly imply
bounds on other physical quantities, e.g., the maximal force exerted
by a molecular motor \cite{HEP}. 
Second, {\em suppose} that (\ref{DD}), and hence (\ref{D>V2}), 
is also valid in {\em infinite} (aperiodic) random systems. Whenever
the diffusion is sublinear, i.e, whenever $\lim_{t\to\infty} (\langle
x^2 \rangle - \langle x \rangle^2)/t = 0$, there is $D=0$. Equation
(\ref{DD}) would then imply that $V=0$, or $\lim_{t\to\infty} \langle
x \rangle/t = 0$. This would mean that whenever diffusion is sublinear
(``subdiffusion'), the drift is also sublinear (``subdrift'') or
vanishes altogether. The asymmetric hopping model with bond disorder
can serve as an example of an infinite system where such relation is
actually observed \cite{Bouchaud}.


\section{Diffusion in 
         arbitrary space dimension \lowercase{$d$}}
\label{SecTotGen}

Suppose we want to calculate the drift velocity $\vec{V}$ and the
diffusion tensor {\bf D} in a system with $d$ Euclidean coordinates.
If the system can be divided into a {\em finite} number of subsystems
with constant transition rates between each two of them, our major
results (\ref{l0}) -- (\ref{h0}) and (\ref{DFPKT1}) remain valid
irrespective of the geometry of the system. Such division is always
possible for periodic systems with finite number of sites in the
elementary cell.  The matrix $\Lambda(\vec{k})$ is the most
``sensitive'' to the geometry of the system under consideration. In
calculating its explicit form one can use an equation similar to
(\ref{DefLambdaK}), remembering, however, that $\vec{k}$ and
$\vec{x}_n$ are now vectors in a $d$-dimensional space. One should
also take into account all possible transitions between sublattices,
and this can be done by considering all possible jumps starting at any
site belonging to some elementary cell and ending in the same cell or
in one of its nearest-neighbor cells.

The components of the velocity vector $\vec{V}$ and the diffusion tensor
${\bf D}$ are given by 
\begin{eqnarray}
  V_\mu &=& 
             \lim_{t\to\infty} \frac{\langle x_\mu \rangle}{t}
          =  \lim_{t\to\infty}
            i\frac{1}{t}
                    \left. \frac{
                                   \partial \tilde{P}(\vec{k},t)
                                }{
                                   \partial k_\mu
                                 }
                    \right|_{\vec{k}=0}, 
    \\[1ex]
  {\bf D}_{\mu\sigma}
         &=&
            \lim_{t\to\infty}
              \frac{ \langle x_\mu x_\sigma \rangle 
                    -\langle x_\mu \rangle \langle x_\sigma \rangle}{2t}
          =          \lim_{t\to\infty} 
              \frac{1}{2t}
 	      \left. 
                  \left(
                    -  \frac{\partial^2\tilde{P}(\vec{k},t)}{
                             \partial k_\mu \partial k_\sigma} 
                    + 
                       \frac{\partial \tilde{P}(\vec{k},t)}{\partial k_\mu}
                       \frac{\partial \tilde{P}(\vec{k},t)}{\partial k_\sigma}
                  \right)
              \right|_{\vec{k}=0}.
\end{eqnarray}
Using (\ref{l0}) -- (\ref{h0}) and (\ref{DFPKT1})  we conclude that
\begin{eqnarray}
  \label{d-DefJ1L}
  V_\mu &=&V_\mu^{\rm D} = 
     i  \left. 
              \frac{\partial \lambda_0}{\partial k_\mu}
         \right|_{\vec{k}=0} , 
        \\[1ex]
  \label{d-DefD1L}
  {\bf D}_{\mu\sigma} &=& 
         \left.
            -\frac{1}{2}
            \frac{\partial^2\lambda_0 }{\partial k_\mu \partial k_\sigma} 
         \right|_{\vec{k}=0} ,
  \\[1ex]
  \label{d-DDefD1L}
  {\bf D}^{\rm D}_{\mu\sigma} &=&  
          \left.
              -\frac{1}{2}
              \left[
                 \frac{\partial^2\lambda^{\rm D}_0}{\partial k_\mu
                                                    \partial k_\sigma}  
                -
                    \frac{\partial\lambda^{\rm D}_0}{\partial k_\mu}
                    \frac{\partial\lambda^{\rm D}_0}{\partial k_\sigma}
              \right] 
          \right|_{\vec{k}=0},
\end{eqnarray}
where $\mu, \sigma = 1,\ldots,d$.
Actually, since $\log\tilde{P}(\vec{k},t) \approx \lambda_0(\vec{k})t$
is a generating function for cumulants of $P(\vec{x},t)$ (see
\cite{vanKampen}), the first two of the above formulae are a natural
consequence of (\ref{FPKT0}) and (\ref{h0}).  

Just as in the one-dimensional case any component of $\vec{V}$ or
${\bf D}$ can be expressed in terms of derivatives of the three lowest
coefficients of the characteristic polynomial of the matrix
$\Lambda(k)$ at $k=0$. In particular,
\begin{eqnarray}
  \label{d-J}
  V_\mu &=& V_\mu^{\rm D} =  
          -i \left.
              \frac{c_0^{(\mu)}}{c_1}
            \right|_{\vec{k}=0},
     \\[1ex]
  \label{d-D}   
  {\bf D}_{\mu\sigma} &=&
     \left.   
      \frac{
              c_0^{(\mu)(\sigma)}
         -2c_2V_\mu V_\sigma 
         - i\left(
               c_1^{(\mu)}    V_\sigma
             + c_1^{(\sigma)} V_\mu
            \right)
           }{2c_1}
     \right|_{\vec{k}=0},
     \\[1ex]
  {\bf D}_{\mu\sigma}^{\rm D} &=&
             {\bf D}_{\mu\sigma}
            -\frac{1}{2}V_\mu V_\sigma .
\end{eqnarray}
where we used a short-hand notation $f^{(\mu)} \equiv \partial f
/\partial k_\mu$.


\section{The nearest-neighbor
         transition rates}
\label{SecNN}

If in a one-dimensional system only transitions to the
nearest-neighbors are allowed, and if the distance between
nearest-neighbor sites is equal 1, our formulae (\ref{SolJ1C}) and
(\ref{SolD1C}) can be simplified by noting that in this particular
case
\begin{eqnarray}
 \label{c_0P}
c_0' &=&  iL\left( 
               \prod_{j=0}^{L-1}\Gamma^{\rightarrow}_j 
             - \prod_{j=0}^{L-1}\Gamma^{\leftarrow}_j
            \right),\\[1ex]
 \label{c_0PP}
c_0'' &=& L^2\left( 
                \prod_{j=0}^{L-1}\Gamma^{\rightarrow}_j 
              + \prod_{j=0}^{L-1}\Gamma^{\leftarrow}_j
             \right),\\[1ex]
 \label{c_1p}
c_1'  &=& 0,
\end{eqnarray}
where $\Gamma^{\rightarrow}_j$ and $\Gamma^{\leftarrow}_j$ are the
transition rates to the right and left from site $j$, respectively;
$\Gamma^{\rightarrow}_j \equiv \Gamma_{j+1,j} $ and
$\Gamma^{\leftarrow}_j \equiv \Gamma_{j-1,j}$.  Thus, to determine $V$
and $D$ we only need to compute the two terms of the characteristic
polynomial of $\Lambda(0)$: $c_1$ and $c_2$.

Although $c_1$ and $c_2$ could be, at least in principle, found by
calculating $\det (x I -\Lambda(0))$, $I$ being the identity matrix,
this could lead to serious computational overhead even for $L$ of
order 10. A more efficient method exploits the fact that $c_l$ are
polynomials in $\Gamma^{\leftarrow}_j$ and $\Gamma^{\rightarrow}_j$,
$j = 0,\ldots,L-1,$ and can be expressed as
\begin{equation}
\label{clPoly}
  c_l = \sum_{\{\gamma_j\},\{\delta_j\}} 
        \;
        \prod_{m, n =0}^{L-1} 
             (\Gamma^{\rightarrow}_m)^{\gamma_m}
             (\Gamma^{\leftarrow}_n)^{\delta_n} 
             \psi_l(\{\gamma_j\},\{\delta_j\}) 
\end{equation}
where $l \in \{1,2\}$, $\gamma_m \in \{0,1\}$, $\delta_n \in \{0,1\}$,
and $\psi_l(\{\gamma_j\},\{\delta_j\}) = 0$ if at least one of the
following conditions is satisfied:
  \begin{eqnarray}
     & \sum_{m=0}^{L-1}(\gamma_m + \delta_m) \neq L - l,& \nonumber\\[1ex]
     & \exists_{m}   \;\;  \gamma_m = \delta_m = 1,&      \nonumber\\[1ex]
     & \exists_{m}   \;\;  \gamma_m = \delta_{m+1} = 1;&
  \end{eqnarray}
otherwise $\psi(\{\gamma_j\},\{\delta_j\}) = 1$. 
For example, for $L = 3$ there is 
\label{L=3}
\begin{eqnarray}
   \label{c1c2L3}
   c_1 &=&
       \PL_2\PL_0 + \PL_2\PR_0 + \PR_2\PR_0    \nonumber  \\[1ex]
     &&+\PL_1\PL_2 + \PL_1\PR_2 + \PR_1\PR_2   \nonumber \\[1ex]
     &&+\PL_0\PL_1 + \PL_0\PR_1 + \PR_0\PR_1 , \nonumber \\[1ex]
   c_2 &=& \PL_0 +\PL_1+ \PL_2+\PR_0 +\PR_1 + \PR_2. 
\end{eqnarray}
Although formally the sum in (\ref{clPoly}) consists of $2^{2L}$ terms,
in practice only for a few of them $\psi_l(\{\gamma_j\},\{\delta_j\})
\neq 0$. In particular, for a given value of $L$ the sum in
(\ref{clPoly}) consist of $L^2$ non-vanishing terms for $c_1$ and
$(L^4-L^2)/12$ terms for $c_2$. 
Thus it should be possible to calculate $D$ algebraically for $L$ of
order 20, and numerically for $L$ of order at least 100. It is worth
noting that since $c_l$ can be computed as sums of positive values, by
using (\ref{clPoly}) rather than calculating the determinant of
$x I - \Lambda(0)$ we can avoid large numerical errors which
often appear when computing determinants of large matrices.

It is not difficult to show that our general formulae (\ref{SolJ1C})
and (\ref{SolD1C}) are consistent with eq.\ (\ref{Kutner}) derived for
a general case of 1D systems at equilibrium with nearest-neighbor
transitions. To this end it suffices to notice that the equilibrium
site occupation probabilities $P^{\rm eq}_j$ can be expressed in terms
of the jump probabilities as
\begin{equation}
 \label{Peq}
 \frac{1}{P^{\rm eq}_j \Gamma^{\rightarrow}_j}
   = \frac{1}{\Gamma^{\rightarrow}_j} 
   + \frac{\Gamma^{\leftarrow}_j}{\Gamma^{\rightarrow}_j
     \Gamma^{\rightarrow}_{j-1}}
   + \ldots
   + \frac{\Gamma^{\leftarrow}_j  \Gamma^{\leftarrow}_{j-1} 
             \ldots \Gamma^{\leftarrow}_{j-L+2}}{
           \Gamma^{\rightarrow}_{j}\Gamma^{\rightarrow}_{j-1}
             \ldots \Gamma^{\rightarrow}_{j-L+1}}.
\end{equation}
On inserting it into (\ref{Kutner}) and then taking into
account (\ref{clPoly}) and the trivial condition $V = 0$ one arrives
at (\ref{SolD1C}).

Another interesting consequence of (\ref{c_0P}) - (\ref{c_1p}) is that
diffusion in systems with transitions to nearest-neighbours is always
normal. To see it note that since $c_1 \neq 0$, $D$ is always finite
and thus no superdiffusion is possible. The other type of anomaly,
subdiffusion, would require that $D=0$. Owing to (\ref{D>V2}) this
would imply $V=0$ which, upon taking into account (\ref{SolD1C}),
would require $c''_0 =0$. However, the explicit form of $c''_0$, see
(\ref{c_0PP}), guarantees that $c''_0 > 0$ except for a degenerated
case where the diffusing particle is confined to a finite region of
the lattice ($\Gamma^{\leftarrow}_{j} = \Gamma^{\rightarrow}_{l} = 0$
for some $j$ and $l$).

One final remark. Some authors \cite{KehrJ,MotorsC,KutnerPHA}, when
considering a one-dimensional system with nearest-neighbor
transitions, prefer to reduce the problem to diffusion on a finite
ring and investigate the probability current $J$ rather than the drift
velocity $V$. The former is defined as $J = P_j\Gamma^{\rightarrow}_j
- P_{j+1}\Gamma^{\leftarrow}_{j+1}$ and in the steady state does not
depend on the site $j$ it is measured at. For single-particle systems
$J$ and $V$ are related to each other through a simple formula $J =
V/L$ \cite{Der83JSP}.  In this case our approach can thus be applied
for calculating both $V$ and $J$.

\section{Applications}
\label{SecAppl}

\subsection{The case $L =1$ and  $L=2$}

Applying our approach to a particle diffusing in a one-dimensional
system with a lattice constant $a=1$, time unit $\tau = 1$, and the
period  $L=1$ we immediately arrive at a well known result
\begin{eqnarray}  
  V &=& \Gamma^{\rightarrow} - \Gamma^{\leftarrow} , \\
   \label{D1}
   D &=& (\Gamma^{\rightarrow} + \Gamma^{\leftarrow})/2, \\
   \label{DD1}
   D^{\rm D} &=& [\Gamma^{\rightarrow} + \Gamma^{\leftarrow} 
                - ( \Gamma^{\rightarrow} - \Gamma^{\leftarrow})^2]/2. 
\end{eqnarray}
For $L = 2$ we find
\begin{eqnarray}
  V &=& 2\frac{  \Gamma^{\rightarrow}_1  \Gamma^{\rightarrow}_2
               - \Gamma^{\leftarrow}_1  \Gamma^{\leftarrow}_2}{S} , \\
   D &=& 2\frac{  \Gamma^{\rightarrow}_1  \Gamma^{\rightarrow}_2 
                + \Gamma^{\leftarrow}_1  \Gamma^{\leftarrow}_2}{S}
         - \frac{V^2}{S}, \\
   D^{\rm D} &=&  2\frac{  \Gamma^{\rightarrow}_1
                           \Gamma^{\rightarrow}_2 
                         + \Gamma^{\leftarrow}_1  
                           \Gamma^{\leftarrow}_2}{S} 
                  -\frac{2+S}{S} \frac{V^2}{2} , 
\end{eqnarray}
where $S \equiv \Gamma^{\rightarrow}_1 + \Gamma^{\rightarrow}_2 +
\Gamma^{\leftarrow}_1 + \Gamma^{\leftarrow}_2$.  Note also that the
solution for $L=3$ can be easily derived using (\ref{c1c2L3}).

\subsection{The sawtooth
            potential of arbitrary period in an external field}

Consider a one-dimensional system of period $L$ with site energies
$E_{nL+j} = j\varepsilon$, where $n$ is an integer, $j=0,\ldots,L-1$,
and $\varepsilon>0$ is a constant. Such a pattern is know as a
discrete sawtooth potential \cite{KehrJ,MotorsB}. We assume that the
transition rate from a site $j$ to $j+1$ is given by
\begin{equation}
\label{SRight}
\Gamma^{\rightarrow}_j = b\exp(-\beta \varepsilon/2),
\end{equation}
and the rate of jumping from  $j$ to $j-1$ reads
\begin{equation}
\label{SLeft}
\Gamma^{\leftarrow}_j = 
   \left\{
      \begin{array}{ll}
         b^{-1}\exp(\beta \varepsilon/2),  
       & j \neq 0\; (\mbox{mod }L) 
       \nonumber \\
        b^{-1}\exp(\beta \varepsilon (1/2-L)),
       & j = 0   \; (\mbox{mod }L)
      \end{array}
   \right.
\end{equation}
where $\beta \equiv 1/k_{\rm B}T$ is the Boltzmann factor and $b\equiv
\exp(\beta F/2)$ represents a bias due to an external force $F$. Note
that with this choice of the transition rates, for $b = 1$ the system
satisfies the detailed balance condition
$\Gamma^{\leftarrow}_{j+1}\exp(-\beta E_{j+1}) =
\Gamma^{\rightarrow}_{j}\exp(-\beta E_{j})$.

Using (\ref{clPoly}) and some combinatorics one can prove that for $L>2$
\begin{mathletters}
 \label{N1Sawtooth}
 \begin{eqnarray}
  \label{N1Sa}
  c_0'  &=& iLR^L(b^{L}-b^{-L})                    \\
  c_0'' &=& L^2R^L(b^{L}+b^{-L})                    \\
  c_1   &=& G^{-L-1}       {\cal S}_{L}(G) 
          + R^{2L}G^{L-1}  {\cal S}_{L-1}(G^{-1})   \\ 
  \label{N1Sd}
  c_1'  &=& 0                                       \\
  \label{conj}
  c_2   &=&  G^{2-L}       {\cal Z}_{L-2}(G) 
           + R^{2L}G^{L-4} {\cal Z}_{L-3}(G^{-1})
 \end{eqnarray}
\end{mathletters}
where $R \equiv \exp(-\beta\varepsilon /2)$ controls the anisotropy of
the potential, $G \equiv Rb$, ${\cal S}_{m}(x) \equiv
\sum_{j=1}^{m}jx^{2j} = [x^{2m+2}(1+m-mx^2) - x^2]/(x^2-1)^2$, and
${\cal Z}_m(x) \equiv \case{1}{2}
\sum_{j=0}^{m}(m+1-j)(j+1)(j+2)x^{2j}$.  Actually we were able to
prove only (\ref{N1Sa})--(\ref{N1Sd}), and (\ref{conj}) is a
conjecture based on the form of $c_2$ derived for $L=2,\ldots,20$.

\begin{figure}
  \epsfxsize6in 
  \epsfbox{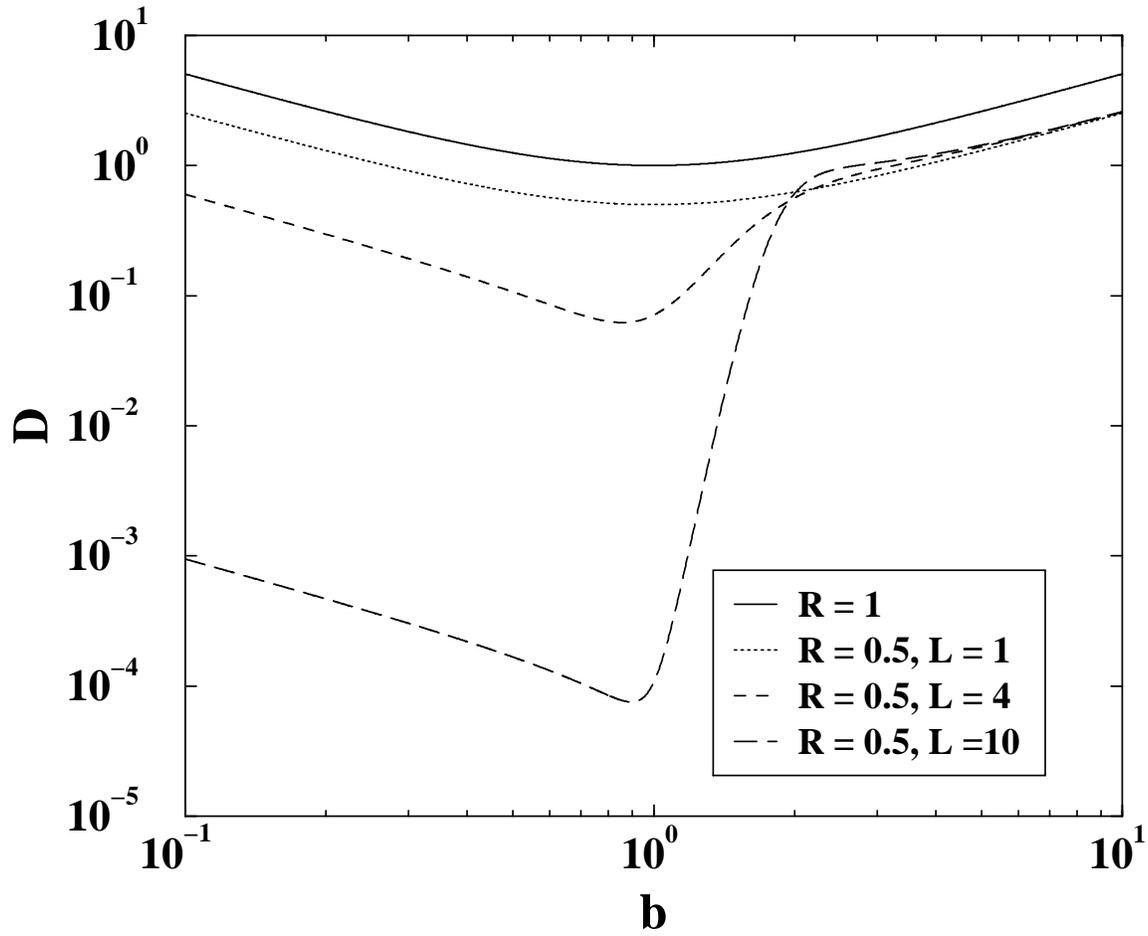}
  \caption{Diffusion coefficient D of a single particle in a sawtooth
  potential (\protect\ref{SRight}) and (\protect\ref{SLeft}) for
  various potential periods $L$ and anisotropy parameters $R$ 
  as a function of the bias $b$. Arbitrary units.  
    \label{Fig1} 
  }
\end{figure}

The stationary drift velocity $V$ and the diffusion constants $D$ and
$D^{\rm D}$ can be now calculated using (\ref{SolJ1C}),
(\ref{SolD1C}), and (\ref{DD}). The resulting velocity has already
been studied in Ref.\ \cite{KehrJ}. The diffusion coefficient $D$
calculated for various values of $R$ and $L$ as a function of the bias
$b$ is depicted in Fig.\ \ref{Fig1}. For $R=1$ there is $\varepsilon
=0$, and so for any $j$ we have $\Gamma^{\rightarrow}_j = b$ and
$\Gamma^{\leftarrow}_j = b^{-1}$. Consequently, the effective period
equals~$1$. Using (\ref{D1}) we conclude that $D =
\case{1}{2}(b+b^{-1})$.  For $R<1$ the behaviour of $D$ becomes more
complicated. For a large bias to the right, $b \gg 1$, the jumps to
the left are so rare that practically they become irrelevant. We thus have
$\Gamma^{\rightarrow}_j = bR$, $\Gamma^{\leftarrow}_j \approx 0$, and
so $D \approx bR/2$ irrespective of $L$. For a strong bias to the
left, $b \ll 1$, the jumps to the right can be neglected, but the
particle has to jump over a large potential barrier located at sites
$j=\ldots,-L,0,L,\ldots$ whose height is proportional to
$L-\case{1}{2}$, see (\ref{SLeft}). Therefore, in this regime $D$ is a
quickly decreasing function of $L$. These two limiting solutions match
in the intermediate regime, which can be roughly described as $1 \le b
\le 1/R$.

\subsection{Two particles
 in a sawtooth potential on a ring of length $L=4$}
\label{2particles}

Consider {\em two} particles diffusing in a sawtooth potential on a
ring of length $L=4$. We assume that any site $j$ can be occupied by
only one of them (the hard-core interaction). For simplicity we also
assume that the distance between the particles cannot exceed $L-1$, so
that we can reduce the system to a ring consisting of $L$ sites.

Each state of the system can be described as a pair of integers,
$(n,m)$, where $n,m=0,\ldots,L-1$ denote the currently occupied
sites. As $n\neq m$, there are $L(L-1)/2 = 6$ different states. Our
two-particle system is thus equivalent to a 6-state system with one
``virtual'' random walker. These states are, in order, $(0,1)$,
$(0,2)$, $(0,3)$, $(1,2)$, $(1,3)$, $(2,3)$.  The distance between a
state $(i_1,j_1)$ and $(i_2,j_2)$ equals $i_2-i_1 + j_2-j_1$. For
example, the distance from $(0,1)$ to $(0,2)$ is 1.  The transition
rate between $(i_1,j_1)$ and $(i_2,j_2)$ vanishes unless $|i_2-i_1| +
|j_2-j_1| = 1$, and in this case a transition from $(i_1,j_1)$ to
$(i_2,j_2)$ corresponds to a single jump of one of two diffusing
particles. For example, the transition rate from $(0,1)$ to $(0,2)$ is
$\Gamma^{\rightarrow}_1$, and from $(0,1)$ to $(0,3)$ is 0.
Consequently, if the lattice constant $a=1$, the matrix $\Lambda(k)$
takes on the following form,
{
 \small
 \begin{equation}
 \Lambda(k) = 
  \left [
   \begin{array}{cccccc}
    -\PR_1 - \PL_0 & \PL_2{\rm e}^{ik} & 0 & 0 & \PR_3 {\rm e}^{-ik} & 0 \\
      \noalign{\medskip}
    \PR_1 {\rm e}^{-ik} & -\PL_2 - \PR_2 - \PR_0 - \PL_0 & \PL_3{\rm e}^{ik} &
      \PL_1 {\rm e}^{ik} & 0 & \PR_3 {\rm e}^{-ik}\\
      \noalign{\medskip} 
    0 & \PR_2 {\rm e}^{-ik} & -\PL_3 -\PR_0 & 0 & \PL_1 {\rm e}^{ik} & 0\\
      \noalign{\medskip} 
    0 & \PR_0 {\rm e}^{-ik} & 0 & -\PL_1 - \PR_2 & \PL_3 {\rm e}^{ik} & 0\\
      \noalign{\medskip}
    \PL_0 {\rm e}^{ik} & 0 & \PR_0 {\rm e}^{-ik} & \PR_2 {\rm e}^{-ik} &
      -\PR_3 - \PL_1 -\PL_3 -\PR_1 & \PL_2 {\rm e}^{ik}\\
      \noalign{\medskip}
    0 & \PL_0 {\rm e}^{ik} & 0 & 0 & \PR_1 {\rm e}^{-ik} & -\PR_3 -\PL_2
   \end {array}
  \right ]
 \end{equation}
}

The drift velocity $V$ and the diffusion coefficient $D$ can be found
for arbitrary transition rates, but the results are {\em very}
lengthy. However, in a particular case of the sawtooth potential the
transition rates $\Gamma^{\rightarrow}_j$ and $\Gamma^{\leftarrow}_j$
are given by relatively simple formulae (\ref{SRight}) and
(\ref{SLeft}), respectively. Then
\begin{eqnarray}
  \label{N2Sawtooth}
  c_0'  &=&  8i G^{-6} (G^8-R^8) (R^8+2G^2+1) (G^2+1) \nonumber \\
  c_0'' &=&  32 G^{-6} (G^8+R^8)  (R^8+2G^2+1) (G^2+1) \nonumber\\
  c_1   &=&  G^{-5}(G^2+1)
            [
             12G^8 + 15G^6 + 10G^4 + 5G^2  + 2  \nonumber\\
        && + R^8    (9G^6 +  13G^4 + 17G^2 + 5) \nonumber\\
        && + R^{16} (G^4 + 2G^2 + 5) 
            ]\nonumber\\ 
  c_1'  &=&  8i G^{-5} (G^8-R^8) (R^8+4G^2+3)  \nonumber\\
  c_2   &=&   G^{-4}[
             37G^8 + 69G^6 + 58G^4 + 31G^2 + 9 \nonumber\\
        && + R^8(23G^6 + 49G^4 + 53G^2 + 19)           \nonumber\\
        && + R^{16}(3G^4 + 8G^2 + 9)                 
            ],
\end{eqnarray}
where $G$ and $R$ were defined below eq.\ (\ref{N1Sawtooth}). 

\begin{figure}
  \epsfxsize6in 
  \epsfbox{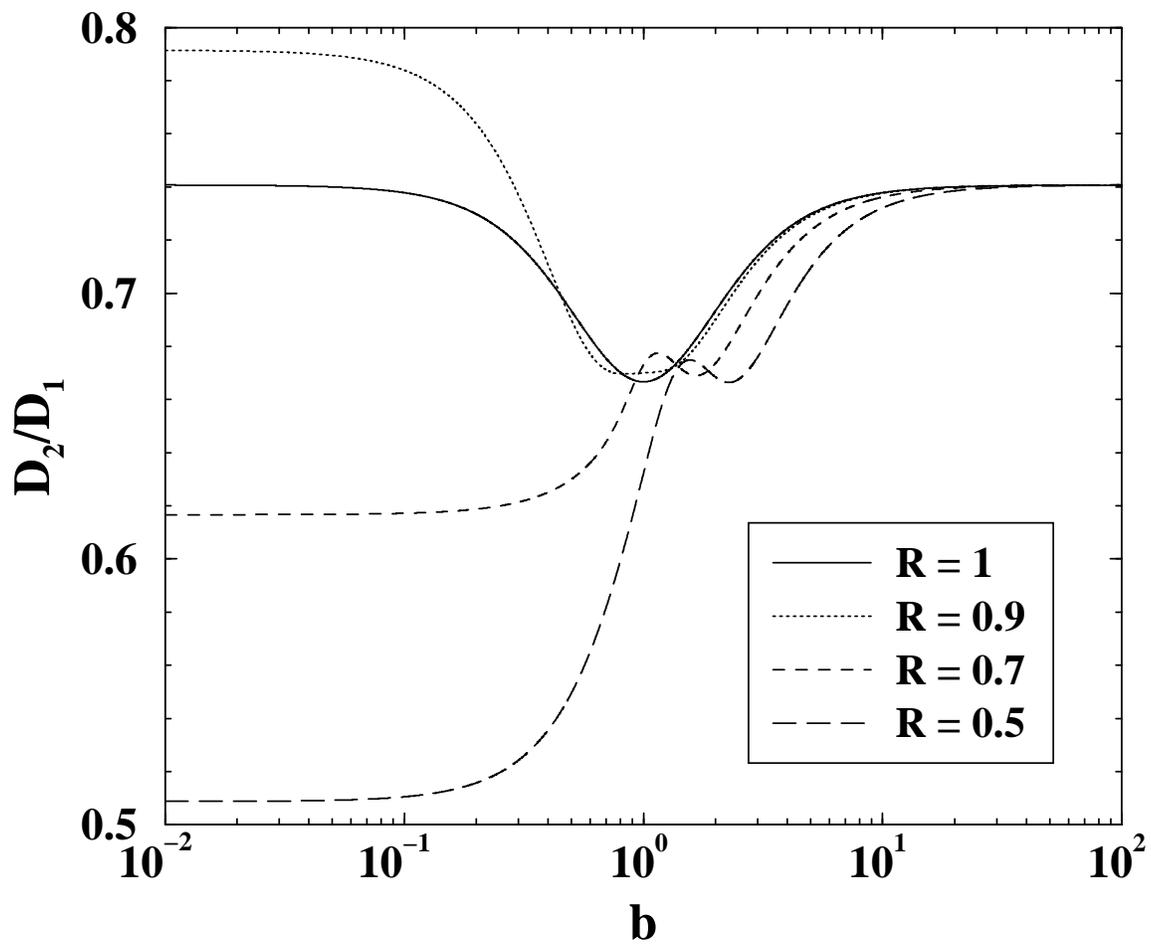}
  \caption{The ratio of the diffusion coefficients $D_2$ and $D_1$
  calculated for a 4-site ring containing 2 and 1 particles, respectively,
  as a function of the bias $b$, for various values of the anisotropy
  parameter  $R$. Arbitrary units.  
    \label{Fig2} 
  }
\end{figure}

Just like in the previous example, $V$, $D$, and $D^{\rm D}$ can be
now determined using (\ref{SolJ1C}), (\ref{SolD1C}), and (\ref{DD}).
The properties of the velocity $V$ will be studied in detail elsewhere
(see Ref.\ \cite{KK}).  Here, in Fig.\ \ref{Fig2} we present the ratio
of the diffusion coefficient calculated for a 2-particle system
($D_2$) to that calculated for a single-particle system ($D_1$), for
$L=4$, as a function of the bias $b$. This ratio measures the change
of diffusivity due to interactions with the second particle.  As could
be expected, $D_2 < D_1$, i.e. the presence of the second particle
decreases the diffusivity of the first one. This effect is not strong,
however, for we find that $D_2 \ge \case{1}{2}D_1$. As in the previous
example, for large bias $b$, the ratio $D_2/D_1$ becomes independent
of the potential anisotropy $R$.


\section{Conclusions}
\label{SecConcl}

We have developed an effective technique of calculating the drift
velocity $\vec{V}$ and the diffusion tensor ${\bf D}$ in arbitrary
periodic systems. A novel feature of our approach, as compared to
previously proposed methods
\cite{Der83JSP,Der83PRB,Lyo,KutnerPHA,Wichmann,Braun}, consists in the
fact that we need not examine in detail the steady-state properties of
the system. In particular, we do not need to solve linear equations of
relatively high order to find the stationary probability distribution
of the random walker over different sublattices.  Instead we calculate
$\vec{V}$ and ${\bf D}$ directly.  The only quantity we really need
in order to carry out our calculations is the matrix $\Lambda(k)$. Its
explicit form, however, can be found almost trivially; all one needs
to know are transition rates and distances from any site of an
elementary cell to any other site in the same or adjacent cell.  Once
the form of $\Lambda(k)$ has been determined, one should calculate the
three lowest terms of its characteristic polynomial. Then $\vec{V}$
and ${\bf D}$ can be determined using our formulae (\ref{d-J}) and
(\ref{d-D}), respectively.

Our approach is very general and can be applied practically to any
periodic system in an arbitrary space dimension.  It is particularly well
suited for calculations employing computer-algebra systems, e.g.,
Maple or Reduce, or for numerical analysis. In this context one should
also note that the coefficients $c_l$ of the characteristic polynomial
of $\Lambda(0)$ can be {\em always} expressed as polynomials in
$\Lambda_{jl}(0)$, $j\neq l$, with {\em positive} coefficients, and
thus can be calculated numerically with extremely high accuracy. While
this is not necessarily true in the case of their derivatives, we
believe that our approach enables one to calculate $\vec{V}$ and
${\bf D}$ even for systems where the corresponding steady-state
problem is numerically ill-conditioned.

We were able to prove a general relation between the diffusion
constants calculated using continuous-time and discrete-time
approaches. Our analysis shows that in {\em any} periodic system
(including, for example, lattice gases studied in
Refs. \cite{Driven,Katz}) they differ by a term $\frac{1}{2} V^2$,
where $V$ denotes the steady-state drift velocity along a given
direction.  Since the limits $t\to\infty$ and $L\to \infty$ were shown
to commute, at least in some simple models \cite{Bouchaud,AslJSP}, we
expect that this relation holds also for infinite random systems.
It would be interesting to know whether such a simple and
apparently universal relation can be applied in contexts other than
those studied here.

We also showed how our technique can be applied to a simple many-body
problem.  While, owing to mathematical complexity, we do not expect
that in this way one will be able to calculate explicitly transport
coefficients in non-equilibrium systems containing more than a few
particles, our conclusion about the universal difference between
diffusion coefficients calculated in continuous- and discrete-time
models should still hold. This conjecture is based on the fact that
diffusion of several particles can be interpreted as diffusion of a
single particle in the corresponding multidimensional phase space, and
for the latter problem the relation between $D$ and $D^{\rm D}$ has
been established rigorously in Section \ref{SecTotGen}.  Thus, for the 
future work, two problems are of primary importance: the role of
periodic boundary conditions in the limit of asymptotically infinite
period, $L\to \infty$, and detailed analysis of possible applications
of our method to many-body systems.

\acknowledgments
I acknowledge stimulating discussions with K.\ W.\ Kehr and H.\ Ambaye.
This work was supported by the Polish KBN Grant
Nr 2 P03B 059 12.


\end{document}